\documentclass[prb,nofootinbib,twocolumn,superscriptaddress]{revtex4} 

\usepackage{graphicx}
\usepackage{dcolumn}
\usepackage{bm}
\usepackage{threeparttable}
\usepackage{times}
\usepackage{mathptmx}
\usepackage{lscape}
\usepackage{natbib}
\usepackage{amsmath}
\usepackage{amssymb}
\usepackage{braket}
\usepackage{comment}

\def\degree{\kern-.2em\r{}\kern-.3em}

\begin{document}


\title{ Analytic Determination of Variance for Configuratinal Density of States in Crystalline Solids }

\author{Tetsuya Taikei}
\affiliation{
Department of Materials Science and Engineering,  Kyoto University, Sakyo, Kyoto 606-8501, Japan\\
}%

\author{Kazuhito Takeuchi}
\affiliation{
Department of Materials Science and Engineering,  Kyoto University, Sakyo, Kyoto 606-8501, Japan\\
}%

\author{Koretaka Yuge}
\affiliation{
Department of Materials Science and Engineering,  Kyoto University, Sakyo, Kyoto 606-8501, Japan\\
}%

\begin{abstract}
{  Our recent study elucidate that information of density of states in configuration space (CDOS) for non-interacting system, characterized by spatial constraint on the system, plays essential role to determine thermodynamically equilibrium properties for interacting systems. Particularly for disordered states, variance of CDOS along all possible coordinations plays significant role. Despite this fact, even for binary system of crystalline solids, analytic expression for variance of CDOS as a function of composition has not been clarified so far. Here we successfully derive variance of CDOS as a function of composition for pair correlations, whose validity is demonstrated by comparing the results with uniform sampling of CDOS on real lattices. The present result certainly advances determining special microscopic state to characterize Helmholtz free energy in classical systems, whose structure is difficult to determine by numerical simulation.  }
\end{abstract}


\maketitle

\section{Introduction}
When the classical system is in equilibrium state, expectation value of physical quantity, including dynamical variables, can be typically obtained through well-known canonical avarage, $Q_r\left(x,T\right) = Z^{-1}\sum_d q_r^{\left(d\right)} \exp\left(-\beta E^{\left(d\right)}\right)$, where summation is taken over all microscopic states at composition $x$ on phase space. For purely-thermodynamic varlables such as free energy, we cannot take ensemble average applied to dynamical variables, since it has a member of entropy. Sicen the number of possible microscopic states considered on phase space astronomically increases when number of constituents increases, it becomes practically difficult to directly estimate, especially free energy, by the definition of partition function. 
Therefore, in order to overcome such problem, several theoretical approaches have been amply developed to effectively sample important microscopic states to estimate dynamical and/oro pure-thermodynamical variables, including Metropolis algorism, entropic sampling and Wang-Landau sampling. 

Although the existing theoretical approaches successfully predict physical quantities in equilibrium state when multibody interactions of the system is once given, they do not focus on whether special microscopic state(s) independent of information about energy and temperature can exist, which characterized such equilibrium properties: In other words, a set of special microscopic state to determine equilibrium properties is unknown \textit{a priori} without information about constituent elements, multibody interactions or temperature, leading to finding a optimal set of such states by existing approaches.

Very recently, we focus on the role of spatial constraint on constituents in classical system (e.g., lattice for crystalline solids, and volume and density for liquids in rigid box), and successfully find a set of special microscopic states, whose structure depends only on the class of spatial constraint and is independent of constituents, interactions and temperature: Information about physical quantity for the special states can well characterize equilibrium properties including free energy, internal energy and elastic modulus. Especially for estimating free energy, corresponding single special state, called Grand Projection (GP) state, should be fundamentally important since it has been considered that in classical system, free energy cannot be described by information about a single microscopic state. 
In our previous study, although condition of structure for GP state is clearly given in mathematical expression, we do not give explicit form of the GP state for practical lattice: This is because structure of GP state can be determined by partial average of configurational density of states (CDOS) along composition, which has not been analytically known even for pair correlation. Especially, the GP state requires information about variance of CDOS along chosen coordination as a function of composition. 

In the present study, in order to overcome such problem, we here focus on A$_x$B$_{1-x}$ binary system on periodic lattice, and provide analytical expression of variance of CDOS for pair correlation as a function of composition, which has not been derived so far. We demonstrate the validity of the derived expression to compare with numerical simulation for binary systems on representative lattices including fcc and bcc.

\section{Derivation and Application}
In the present study, to describe pair correlation for given atomic configuration $\vec{\sigma}$on lattice, we employ generalized Ising model (GIM), whose basis function can completely specify any possible atomic arrangements. Here, $\sigma_i$ corresponds to spin variable to specify occupation of lattice site $i$ by A of $\sigma_i = +1$ and B of $\sigma_i = -1$.  
In the GIM, correlation function for pair figure $m$ is given by
\begin{eqnarray}
\label{eq:p}
\xi_m \left(\vec{\sigma}\right) = \Braket{\sigma_i \sigma_k}_{m, \textrm{lattice}}, 
\end{eqnarray}
where $\Braket{\cdot}_{m, \textrm{lattice}}$ denotes taking arithmetic average over all pair site $i$ and $k$ forms pair $m$. 
In order to consider variance of CDOS along pair $m$, we rewrite Eq.~(\ref{eq:p}) in more explicit form:
\begin{eqnarray}
\xi_m \left(\vec{\sigma}\right) = \left(2 D N\right)^{-1}  \sum_{i, k} g_m\left(i,k\right) \sigma_i\left(\vec{\sigma}\right) \sigma_k\left(\vec{\sigma}\right), 
\end{eqnarray}
where $N$ and $D$ respectively denotes number of lattice points and number of pair $m$ per site, $g_m\left(i,k\right) = 1$ if site $i$ and $k$ forms pair $m$ and $g_m\left(i,k\right) = 0$ for otherwise, and summation is taken over all lattice points in the system. Note that we here consider that all lattice point for empty lattice is symmetry-equivalent, which holds for representative lattices including fcc, bcc and diamond. Using the above equation, variance of CDOS along pair $m$ can be explicitly given by
\begin{widetext}
\begin{eqnarray}
\label{eq:xi}
\Braket{\xi_m}_{\textrm{var}} &=& \Braket{\xi_m^2}_{\vec{\sigma}} - \Braket{\xi_m}_{\vec{\sigma}}^2 =  \left\{ \left(4D^2 N^2\right)^{-1} \sum_{i,k}\sum_{p,q} g_m\left(i,k\right) g_m\left(p,q\right) \Braket{ \sigma_i \sigma_k \sigma_p \sigma_q }_{\vec{\sigma}} \right \} - \left(2x-1\right)^4,
\end{eqnarray}
\end{widetext}
where $\Braket{\cdot}_{\vec{\sigma}}$ denotes taking average over all possible atomic arrangement at composition $x$ considered. The last equation can be obtained since $g_{m}\left(r,s\right)$ is independent of atomic arrangement and it has been shown that $\Braket{\xi_m}_{\vec{\sigma}} = \left(2x-1\right)^2$ for any pair $m$.  
Since exact expression of $\Braket{\xi_m}_{\textrm{var}}$ is expected to have significantly complex function of both $N$ and $x$, we here focus on capturing global landscape of $\Braket{\xi_m}_{\textrm{var}}$ for sufficiently large $N$. 
From Eq.~(\ref{eq:xi}), we can see that when $\Braket{\xi_m}_{\textrm{var}}$ is expressed as a continuous function of composition $x$, it is reasonably given by polynomial of up to degree 4, since $\Braket{\xi_m^2}_{\vec{\sigma}} $ is given by up to four site correlation. Under this condition, we can start from expressing variance of CDOS along pair $m$ as 
\begin{eqnarray}
\Braket{\xi_m}_{\textrm{var}}\left( x \right) = f\left( x \right) = \sum_{i=0}^4 a_i x^i, 
\end{eqnarray}
where coefficients $a_i$s can depend on $N$ and $D$.  Since we have five unknown quantities $a_0,\cdots, a_4$, we should require five independent conditions for $\Braket{\xi_m}_{\textrm{var}} $. Since at $x=0$ and $x=1$, we have respectively a single atomic arrangement (i.e., pure A and B), we get
\begin{eqnarray}
\label{eq:c1}
f\left(  0 \right) = f\left(  1 \right) = 0. 
\end{eqnarray}
Since pair correlation, $\Braket{\sigma_i \sigma_k}_{m,\textrm{lattice}}$, is symmetric by exchanging all A and B constituents, $\Braket{\xi_m}_{\textrm{var}}\left( x \right)$ is symmetric with respect to $x=1/2$. Applying this condition, we can obtain
\begin{eqnarray}
\label{eq:c2}
\left.\frac{df}{dx} \right|_{x=1/2} &=& 0 \nonumber \\
\left.\frac{df}{dx}\right|_{x=0} &=& \left.\frac{df}{dx}\right|_{x=1}.
\end{eqnarray}
From Eqs.~(\ref{eq:c1}) and~(\ref{eq:c2}), we have obtained four independent conditions, and therefore, we should require one remaining condition. Eqs.~(\ref{eq:c1}) and~(\ref{eq:c2}) do not contain any information about $N$ or $D$, which naturally indicates that the remaining condition should be obtained by explicitly consider Eq.~(\ref{eq:xi}). 
In the present study, we assume that at $N\to\infty$, configurational average of four spin product, $\Braket{ \sigma_i \sigma_k \sigma_p \sigma_q }_{\vec{\sigma}}$, is treated by taking product over independently occupied spin variables depending only on composition $x$. 
This would be allowed when we consider atomic arrangements over \textit{whole} composition, since under this condition, we can independently give spin variable of +1 or -1 at each lattice site to construct IRC. However, we here consider that $\Braket{\xi_m}_{\textrm{var}}$ at \textit{fixed} composition. 
Therefore, in order to apply our assumption for the four spin product, the condition
\begin{eqnarray}
\label{eq:lim}
\lim_{N\to\infty}\Braket{\xi_m}_{\textrm{var}} \left( x \right) = \lim_{N\to\infty}\overline{\Braket{\xi_m}}_{\textrm{var}}
\end{eqnarray}
should be satisfied, where $\overline{\Braket{\xi_m}}_{\textrm{var}}$ denotes variance of CDOS along pair $m$ considering whole composition simultaneously. 
It is clear that the above condition holds true only at equiatomic composition, $x=1/2$. 
Using these prerequisites, we can give explicit expression for the first term of right-hand side of Eq.~(\ref{eq:xi}): 
\begin{eqnarray}
\label{eq:4s}
&&\sum_{i,k}\sum_{p,q} g_m\left(i,k\right) g_m\left(p,q\right) \Braket{ \sigma_i \sigma_k \sigma_p \sigma_q }_{\vec{\sigma}} \nonumber \\
&&= 2\sum_{i,k} g_m\left( i,k \right)^2 \Braket{\sigma_i^2\sigma_k^2}_{\vec{\sigma}}  \nonumber \\
&& + \sum_{\substack{i,k=p,q \\ i\neq q}}  g_m\left( i,k \right) g_m\left( k,q \right) \Braket{\sigma_i \sigma_k^2 \sigma_q}_{\vec{\sigma}} \nonumber \\
&& + \sum_{\substack{i,k,p,q \\ i\neq p,q \\ k\neq p,q}} g_m\left( i,k \right) g_m\left( p,q \right)\Braket{\sigma_i\sigma_k\sigma_p\sigma_q} \nonumber \\
&& = 2DN\cdot2 + 2DN\cdot 2\left( 2D - 1 \right) \left( 2x-1 \right)^2 \nonumber \\
&&+ 2DN\cdot\left\{ 2DN - 2\left( 2D - 1 \right) - 2 \right\}\left( 2x-1 \right)^4.
\end{eqnarray}
In the last equation, factor 2 in the first term comes from possible permutations of $i=p,k=q$ and $i=q, k=p$. 
When we substitute Eq.~(\ref{eq:4s}) into Eq.~(\ref{eq:xi}), we can obtain variance of CDOS at $x=1/2$ as 
\begin{eqnarray}
\label{eq:c3}
\Braket{\xi_m}_{\textrm{var}}\left( \frac{1}{2} \right) = \left( DN \right)^{-1}. 
\end{eqnarray}
Note that the form of Eq.~(\ref{eq:c3}) is the same as variance of pair correlation for random alloy at equiatomic composition previously shown,\cite{ra} although the prerequisites and explicit expression for four spin product, including Eqs.~(\ref{eq:lim}) and~(\ref{eq:4s}) in the previous study remains unclear. 

Using Eqs.~(\ref{eq:c1}), (\ref{eq:c2}) and (\ref{eq:c3}), we can finally obtain five independent conditions, and thus, we can provide analytic expression for variance of CDOS as
\begin{eqnarray}
\Braket{\xi_m}_{\textrm{var}}\left( x \right) = \frac{1}{DN}\left( 16x^4 - 32x^3 + 16x^2 \right),
\end{eqnarray}
which means that corresponding standard deviation is given by simple function of
\begin{eqnarray}
\label{eq:sd}
\Braket{\xi_m}_{\textrm{sd}}\left( x \right) = \frac{4x\left( 1-x \right)}{\sqrt{DN}}.
\end{eqnarray}

\begin{figure}[h]
\begin{center}
\includegraphics[width=1.02\linewidth]{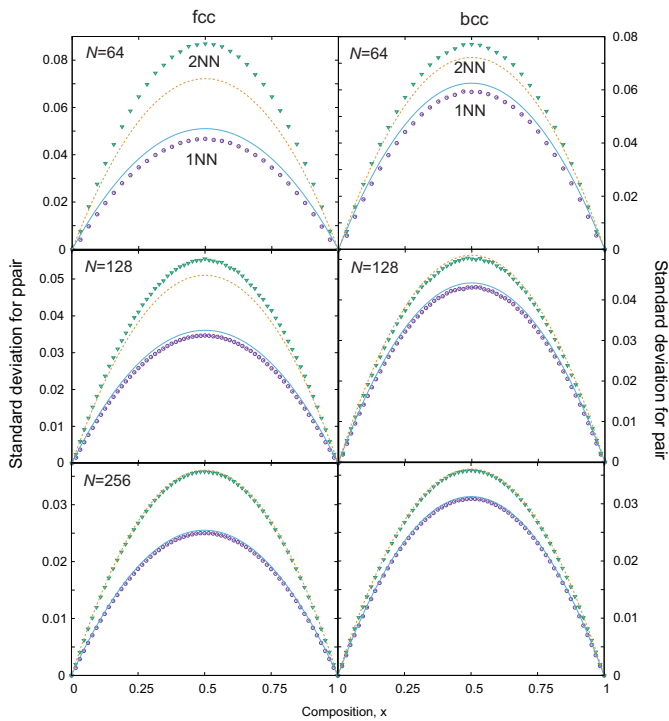}
\caption{ Standard deviation of CDOS along nearest neighbor pair on fcc (left) and bcc (right) lattice, as a function of composition $x$ and number of atoms, $N$. Open circle and triangles denote results of numerical simulation for 1NN and 2NN pair, and solid and broken curves denote those of derived expression, Eq.~(\ref{eq:sd}), for 1NN and 2NN pair. }
\label{fig:sd}
\end{center}
\end{figure}

In order to demonstrate the validity of derived standard deviation, Eq.~(\ref{eq:sd}), we perform Monte Carlo (MC) simulation uniformly sampling possible atomic arrangements for 64-, 128- and 256-atom binary system on fcc and bcc lattices along 1NN and 2NN pairs, to obtain $x$ and $N$ dependence of $\Braket{\xi_m}_{\textrm{sd}}$ for practical lattices. 
Figure~\ref{fig:sd} shows the resultant $\Braket{\xi_m}_{\textrm{sd}}$ for numerial simulation compared with the derived analytic expression of Eq.~(\ref{eq:sd}). Here, $D$ for fcc 1NN, 2NN, and bcc 1NN and 2NN pair respectively takes 6, 3, 4 and 3. 

We can clearly see from Fig.~\ref{fig:sd} that when number of atoms in the system increases, standard deviation for CDOS along chosen pair by numerical simulation unversally agree with our derived expression. 
Particularly at $N=256$, standard deviation given by Eq.~(\ref{eq:sd}) exhibits excellent agreement with numerical simulation: We also confirm that when we further increase number of atoms in the system, deviation between Eq.~(\ref{eq:sd}) and numerical simulation is within accuracy of the simulation. Thus, validity of Eq.~(\ref{eq:sd}) is now well-demonstrated. Additional important point is that 
asymptotic behavior of standard deviation to Eq.~(\ref{eq:sd}) does not always monotonic: For instance, while $\Braket{\xi_m}_{\textrm{sd}}$ for 1NN pair on fcc in numerical simulation exhibit negative deviation from Eq.~(\ref{eq:sd}) and deviation decreases with increase of $N$, 2NN pair on bcc exhibits both positive and negative deviation depending on $N$. 

Finally, we note that the present results of analytic expression of $\Braket{\xi_m}_{\textrm{sd}}$ for pairs certainly advance determination of a special microscopic state, which we call "Grand Projection state (GP state)", whose physical quantity characterize free energy landscape over whole composition in thermodynamically equilibrium state of classical systems. To construct GP state on given lattice, we should know paritial average of $\Braket{\xi_m}$ for possible pairs over multiple compositions of $x\ge 0.5$ and/or $x\le 0.5$ for binary system. However, concept of the GP state relies on the CDOS at sufficiently large $N$ where CDOS along chosen coordination is well-smoothed near equiatomic composition, which directly means that treatment of partial average \textit{at} the equiatomic composition is non-trivial for numerical simulation. Therefore, analytic determination of CDOS (especially, up to 2nd-order moment, corresponding to $\Braket{\xi_m}_{\textrm{sd}}$ we considered here) as a function of composition, Eq.~(\ref{eq:sd}), is significantly useful to construct the GP state without performing undesirable numerical simulation. 

\section{Conclusions}
For crystalline solids, we derive analytic expression of variance for configurational density of states (CDOS) along chosen pair, as a function of composition. We demonstrate the validity of the derived expression by comparing the results for nearest and second-nearest neighbor pairs on fcc and bcc lattices, where we find successful agreement between the expression and simulation with increase of system size. The present result can be significantly useful to determine special microscopic state to characterize free energy landscape, which we recently find for classical systems, without performing undesired numerical simulation.

\section*{Acknowledgement}
This work was supported by a Grant-in-Aid for Scientific Research (16K06704) from the MEXT of Japan, Research Grant from Hitachi Metals$\cdot$Materials Science Foundation, and Advanced Low Carbon Technology Research and Development Program of the Japan Science and Technology Agency (JST).

\end{document}